# Estimate of the Critical Exponents from the Field-Theoretical Renormalization Group: Mathematical Sense of the "Standard Values"

A. A. Pogorelov and I. M. Suslov

*Kapitza Institute for Physical Problems, Russian Academy of Sciences, ul. Kosygina 2, Moscow, 119334 Russia*
*e-mail: suslov@kapitza.ras.ru*

**Abstract**—New estimates of the critical exponents have been obtained from the field-theoretical renormalization group using a new method for summing divergent series. The results almost coincide with the central values obtained by Le Guillou and Zinn-Justin (the so-called "standard values"), but have lower uncertainty. It has been shown that usual field-theoretical estimates implicitly imply the smoothness of the coefficient functions. The last assumption is open for discussion in view of the existence of the oscillating contribution to the coefficient functions. The appropriate interpretation of the last contribution is necessary both for the estimation of the systematic errors of the "standard values" and for a further increase in accuracy.

PACS numbers: 11.10.Kk, 11.15.Pg, 11.15.Me, 64.60.F-, 75.10.Hk

## 1. INTRODUCTION

The most justified theoretical approach to the calculation of the critical exponents is the field-theoretical renormalization group approach [1]. It is based on the description of phase transitions by the effective $\varphi^4$ theory with the action

$$S\{\varphi\} = \int d^d x \left\{ \frac{1}{2}(\nabla\varphi)^2 + \frac{1}{2}m^2\varphi^2 + \frac{1}{4}g\varphi^4 \right\}, \quad (1)$$

where $\varphi$ is the $n$-component field vector, $d$ is the space dimension, and $g$ is the coupling constant. The renormalizability condition of the theory is expressed by the Callan–Symanzik equation

$$\left[ \frac{\partial}{\partial \ln m} + \beta(g)\frac{\partial}{\partial g} + \left(L - \frac{N}{2}\right)\eta(g) - L\eta_2(g) \right] \times \Gamma^{L,N} = 0, \quad (2)$$

where $\Gamma^{L,N}$ is the vertex with $N$ external lines of the field $\varphi$ and $L$ external interaction lines and $\beta(g)$, $\eta(g)$, and $\eta_2(g)$ are the renormalization group functions. The stationary point $g^*$ of the renormalization group is determined as a nontrivial root of the equation $\beta(g) = 0$; then, the critical exponents $\eta$ and $\nu$, as well as the exponent $\omega$ of the correction to scaling, are determined by the expressions

$$\eta = \eta(g^*), \quad \nu^{-1} = 2 - \eta(g^*) + \eta_2(g^*),$$
$$\omega = \beta'(g^*), \quad (3)$$

and other exponents are expressed in terms of these three exponents by means of the known relations [2]. The renormalization group functions $\beta(g)$, $\eta(g)$, and $\eta_2(g)$ are given by power series in $g$ for which several first expansion coefficients and high-order asymptotic expression, which is calculated by the Lipatov method [3, 4], are known. Since the series are factorially divergent, their summation requires special methods.

The aim of this investigation was to significantly increase the accuracy of the determination of the critical exponents as compared to classical works [5–7] with the use a new method for summing divergent series [8]. We actually has succeeded to reduce the error of the results and, most important, to achieve the transparent estimate of this error. Nevertheless, optimistic predictions of the increase in the accuracy [8, 9] were not justified, because fundamental difficulties discussed below have been revealed.

The paper is organized as follows. Comparative analysis of the existing methods for summing divergent series is given in Section 2, where the principle possibility of increasing accuracy is also justified. Section 3 presents the results of "natural" summation, which almost coincide with the central values obtained in [6] (the so-called standard values), but have smaller error. Comparison with other sources of information on the critical exponents is performed in Section 4. Section 5 is devoted to the discussion of the problem of the oscillating contribution to the coefficient functions, which prevents a further increase in the accuracy and can be a source of systematic errors inherent in the standard values.

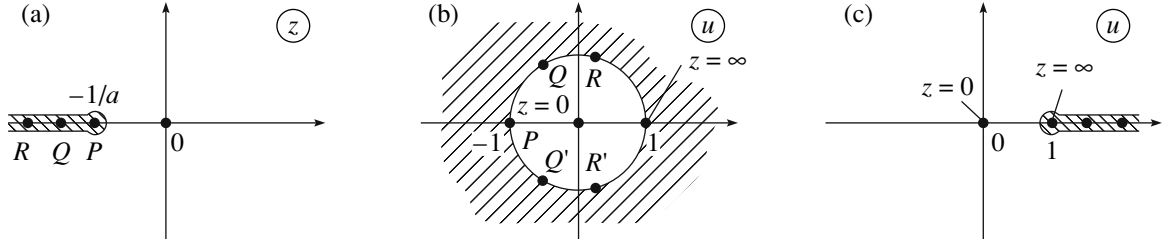

**Fig. 1.** (a) Borel transform $B(z)$ is analytic in the complex plane with the $(-\infty, -1/a)$ cut and (b) its domain of analyticity can be conformally mapped onto the unit circle. (c) The modified conformal transformation on the plane with the $(1, \infty)$ cut can be used for analytic continuation on the positive semiaxis.

## 2. METHODS FOR SUMMING SERIES FOR RENORMALIZATION GROUP FUNCTIONS

Let us consider the power series

$$W(g) = \sum_{N=N_0}^{\infty} W_N(-g)^N \quad (4)$$

whose coefficients have the factorial asymptotic behavior $ca^N\Gamma(N+b)$, which is a usual result of the application of the Lipatov method [9]. The Borel transformation

$$W(g) = \int_0^{\infty} dx\, e^{-x} x^{b_0-1} B(gx),$$

$$B(z) = \sum_{N=N_0}^{\infty} B_N(-z)^N, \quad (5)$$

$$B_N = \frac{W_N}{\Gamma(N+b_0)},$$

where $b_0$ is an arbitrary parameter, reduces the problem to the determination of the Borel transform $B(z)$, which is analytic in the complex $z$ plane with a cut from $-1/a$ to $-\infty$ [10] (see Fig. 1a). The series for $B(z)$ converges in the circle $|z| < 1/a$ and integration in Eqs. (5) requires its analytic continuation beyond the circle. The method for solving this problem determines the basic difference between the methods discussed below.

### 2.1. Padé–Borel Method

The simplest method for analytic continuation is based on the construction of Padé approximants [5]: the function $B(z)$ is approximated by the ratio $P_M(z)/Q_L(z)$ of the polynomials of the degrees $M$ and $L$ with the coefficients chosen so that the known first terms of expansion (5) are reproduced. It is known that the diagonal ($M = L$) and quasi-diagonal ($M \approx L$) approximants for a wide class of functions converge to the desired function in the limit $M \longrightarrow \infty$, which ensures the application of the method. However, for finite $M$ and $L$ values, it reduces to an arbitrary extrapolation in the strong coupling region. Indeed, the use of the Padé approximation imposes the certain (and generally incorrect) behavior of the Borel transform at infinity, $B(z) \sim z^{M-L}$. A similar behavior $W(g) \sim g^{M-L}$ is obtained for the function $W(g)$ in the $g \longrightarrow \infty$ limit, which induces a significant error at $g \sim 1$ due to continuity. In fact, the results of the application of the method are satisfactory, because the real expansion parameter $ag$ near the fixed point is rather small,

$$ag^* \approx 0.2 \ (d=3), \quad ag^* \approx 0.4 \ (d=2), \quad (6)$$

so that RG functions in this region are quite reliably determined by the first terms of the expansion. Nevertheless, the results obtained with different choices of the Padé approximants are significantly different, and the method is used with a certain subjectivity.

The situation is different if at least a rough estimate can be obtained for the asymptotic behavior of the function $W(g)$ in the strong-coupling region. In this case, the behavior of the Padé approximants at infinity can be matched with the asymptotic behavior of the function $B(z)$, and the accuracy uniform in $g$ can be obtained (e.g., if $W(g) \sim g^\alpha$, the approximants with $M - L \approx \alpha$ should be used). If a sufficiently long expansion exists, the strong-coupling asymptotic behavior can be probed using the convergence rate as it was done in the classical paper by Baker et al. [5].

If the Padé approximants are chosen so that all their poles are located on the negative semiaxis, the Padé–Borel method allows one to take into account the analytic properties of the function $B(z)$. Another advantage of this method is the possibility of the direct use of the information on high orders: if transformation (5) is made with $b_0 = b$, the singularity at the point $-1/a$ is a simple pole the residue at which is determined by the parameter $c$ of the Lipatov asymptotics. This circumstance can be used in construction of the Padé approximants.[1] In the last case, the Padé approximation automatically produces interpolation of the coefficient fun-ction (probably, rather smooth); its character has never been controlled, but this is possible in principle.

---

[1] The complete implementation of this program was never performed. Only the position and character of the singularity (i.e., the parameters $a$ and $b$) were taken into account in [5], whereas the residue at it (the parameter $c$) was disregarded.

The characteristic properties of the Padé–Borel method can be summarized as follows:

(i) an arbitrary character of the extrapolation to the strong-coupling region,

(ii) the possibility of the direct inclusion of the Lipatov asymptotic behavior,

(iii) the possibility of taking into account the analytic properties of the function $B(z)$,

(iv) the automatic interpolation of the coefficient function when (ii) is satisfied.

### 2.2. Conformal Borel Technique

The actual analytic continuation of the Borel transform $B(z)$ with numerically specified coefficients $W_N$ is a problem. This problem was elegantly solved by Le Guillou and Zinn-Justin [6] using the conformal transformation $z = f(u)$ of the plane with the cut onto the unit circle $|u| < 1$ (see Fig. 1b). In this case, the re-expansion of the function $B(z)$ in powers of $u$,

$$B(z) = \sum_{N=N_0}^{\infty} B_N(-z)^N \Big|_{z=f(u)} \longrightarrow B(u)$$
$$= \sum_{N=N_0}^{\infty} U_N u^N, \qquad (7)$$

gives a convergent series at any $z$ values. Indeed, all possible singularities $P, Q, R, \ldots$ of the function $B(z)$ lie on the cut, whereas their images $P, Q, Q', R, R', \ldots$ lie at the boundary of the circle $|u| = 1$, so that the second series in Eqs. (7) converges at all values $|u| < 1$, but the interior of the circle $|u| < 1$ is in one-to-one correspondence with the analyticity region in the $z$ plane.

It is easy to see that the main demerit of the Padé–Borel method is overcome in this approach: the analytic continuation procedure is completely objective and is not associated with any arbitrariness. However, difficulties with the use of all available information exist. The knowledge of the first $L$ coefficients of series (4) allows one to determine the same number of the first coefficients $U_N$ in the second series in Eqs. (7); the Lipatov asymptotic behavior is not used in the explicit form and only its parameter $a$ is used to construct the conformal transformation $z = f(u)$. In fact, even parameter $a$ is not used! Indeed, the cut in the Borel plane can be drawn from $-\infty$ to an arbitrary point $z^*$ such that $-1/a < z^* < 0$. In this case, all singularities of the Borel transform are also at the boundary of the unit circle in the $u$ plane and the resumed series converges at all points of the $z$ Borel plane lying beyond the cut. Under this condition, the results are independent of $z^*$. This independence was empirically found in [11] and means that it is unnecessary to know the exact $a$ value; the latter is used only as the rough estimate, determining the bound of the allowable interval for $z^*$. Thus, the direct use of the Lipatov asymptotic behavior appears to be impossible.

It was indirectly taken into account in [6] by interpolating the coefficient function, which made it possible to predict and to use one or two unknown coefficients $W_N$. The use of a large number of the coefficients is impossible owing to the catastrophic increase in the errors. If $\delta$ is the relative accuracy of the calculation of $W_N$, the error in the coefficients $U_N$ of the resumed series increases as [8][2]

$$\delta U_N \sim 5.8^N \delta, \qquad (8)$$

already the third predicted coefficient $W_N$ is inefficient at an interpolation accuracy of 1%. The impossibility of the control over the intermediate coefficients of the expansion leads to the strong dependence of the results on the summation procedure. This uncertainty cannot be analyzed and it is limited using semi-empirical receipts.

The characteristic properties of the conformal-Borel procedure are as follows:

(i) the strictly justified method for analytic continuation,

(ii) the automatic inclusion of the analytic properties of the function $B(z)$,

(iii) the impossibility of the direct use of the Lipatov asymptotic behavior,

(iv) the limited efficiency of the interpolation of the coefficient functions.

### 2.3. Variational Perturbation Theory

The variational perturbation theory whose most efficient variant was proposed by Kleinert [12] belongs to a different kind of methods. The equivalent expansion of the renormalized charge $g$ in the powers of the bare charge $g_0$ (with a specific their definition) is used instead of the expansion in the series for the $\beta$ function. In this case, $g \approx g_0$ and $g \longrightarrow g^*$ for the weak- and strong-coupling regions $g_0 \longrightarrow 0$ and $g_0 \longrightarrow \infty$, respectively. Similar relations are valid for the renormalization group functions, e.g., $\eta(g) \sim g_0^2$ for $g_0 \longrightarrow 0$ and $\eta(g) \longrightarrow \eta(g^*)$ for $g_0 \longrightarrow \infty$. Interpolation between these regions is similar to the use of the Padé–Borel method with the known (constant) strong coupling asymptotic behavior: the diagonal approximants $P_M(g_0)/Q_M(g_0)$ constructed by using the first terms of the series converge well with an increase in $M$, predicting the $g^*$, $\eta(g^*)$, etc. Another set of the approximating functions, which takes into account the character of the approaching $g$ to the fixed point, $g - g^* \sim g_0^{-\omega/(4-d)}$, is used in the Kleinert variant.

---

[2] Strictly speaking, this result was obtained in [8] for random errors. Practice shows, that the result is approximately the same for smooth errors.

Neither the divergence of the series nor its Lipatov asymptotic behavior is used in this approach. Information on the Lipatov asymptotic behavior can be taken into account only indirectly by interpolating the coefficient function (similar to Subsection 2.2) and is not too efficient. This approach has no one of the indisputable advantages inherent in two above methods and is based only on the quality of the interpolation scheme, which is really high. Nevertheless, this scheme has no deep sense and allows subjective variations.

### 2.4. Modified Conformal Borel Technique

Let us describe an algorithm that will be used below. It is based on the idea used in Subsection 2.2, but involves the conformal transformation onto the plane with the $(1, \infty)$ cut (see Fig. 1c) rather than onto the unit circle,

$$z = \frac{u}{a(1-u)}, \qquad (9)$$

for which it is easy to find the relation between $U_N$ and $B_N$:

$$U_0 = B_0,$$

$$U_N = \sum_{K=1}^{N} \frac{B_K}{a^K}(-1)^K C_{N-1}^{K-1} \quad (N \geq 1), \qquad (10)$$

where $C_N^K = N!/K!(N-K)!$ are the binomial coefficients. The series for $B(u)$ converges for $|u| < 1$ and, in particular, on the interval $0 < u < 1$, which is the image of the positive semiaxis. This is sufficient for integration in Eqs. (5). Conformal mapping (9) is convenient for investigating the function $W(g)$ in the strong-coupling region [8], because the convergence of the re-expanded series in Eqs. (7) is determined by the nearest singularity $u = 1$ associated with the singularity of the function $W(g)$ at $g \longrightarrow \infty$. For this reason, the asymptotic behavior of $U_N$ for large $N$ values,

$$U_N = U_\infty N^{\alpha-1}, \quad U_\infty = \frac{W_\infty}{a^\alpha \Gamma(\alpha)\Gamma(b_0+\alpha)}, \qquad (11)$$

is connected with the behavior of the function $W(g)$ for large $g$ values:

$$W(g) = W_\infty g^\alpha, \quad g \longrightarrow \infty. \qquad (12)$$

This algorithm ensures the much slower increase in random errors (e.g. calculational or round-off errors):

$$\delta U_N \sim \delta \cdot 2^N \qquad (13)$$

(cf. Eq. (8)). The quantity $\delta U_N$ in the double-precision computer calculations ($\delta \sim 10^{-14}$) becomes on the order of unity at $N \approx 45$, which allows the reconstruction of asymptotic behavior (12) at a level of about 1%. Then, the coefficients $U_N$ are continued to the large $N$ values according to found power law (11). Thus, all coefficients of converging series (7) are known, and this series can be summed at arbitrary $g$ values.

The effect of smooth errors is even more interesting, and the algorithm is characterized by a peculiar superstability. Linear transformation (10) has the remarkable property

$$\sum_{K=1}^{N} K^m (-1)^K C_{N-1}^{K-1} = 0 \qquad (14)$$

for $m = 0, 1, \ldots, N-2,$

and the addition of an arbitrary polynomial $P_m(K)$ to $B_K/a^K$ (having a power-law behavior in $K$) does not affect the large-$N$ behavior of $U_N$. A similar property is inherent in a wide class of smooth functions that are well approximated by polynomials. In particular, change in $U_N$ under the change $B_K/a^K \longrightarrow B_K/a^K + f(K)$, where $f(K)$ is an entire function with the rapidly decreasing coefficients of the Taylor series, decreases rapidly with an increase in $N$. For the renormalization group functions, first several coefficients $W_N$ and their asymptotic behavior are known, and the intermediate coefficients can be determined by interpolation. Interpolation errors are smooth and appear to be insignificant even when they are large.

As a result, the coefficient function can be interpolated, the resulting series can be unambiguously summed, and the uncertainty of the results can be analyzed by varying the interpolation procedure. Thus, the problem of the dependence of the results on the summation procedure is completely eliminated, and only the dependence of the results on the interpolation method remains. The latter has a direct physical sense and is associated with the incompleteness of the initial information.

In contrast to the Padé–Borel method (see Subsection 2.1), the described algorithm involves the explicit estimate of the strong-coupling asymptotic behavior. In contrast to the standard conform-Borel procedure (see Subsection 2.2) and variational perturbation theory (see Subsection 2.3), it ensures the complete inclusion of available information. Thus, one can hope to increase the accuracy already with available information.

### 3. SUMMATION RESULTS

The application of the method begins with the interpolation of the coefficient function by using the formula[3]

---

[3] Corrections to the Lipatov asymptotic behavior have the form of a regular expansion in $1/N$, which, after the re-expansion, leads to formula (15).

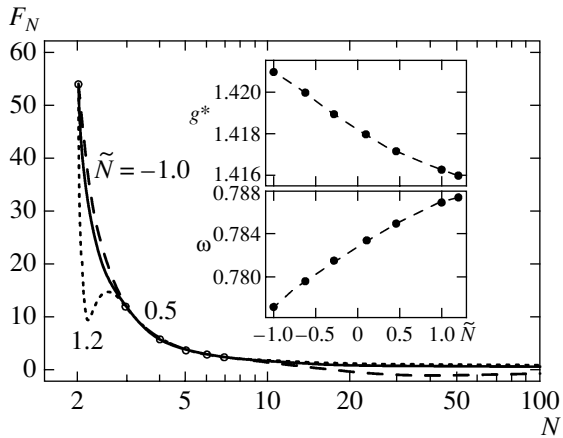

**Fig. 2.** Interpolation curves for the function $\beta(g)$ at $L_0 = 2$. The insets show the results for $g^*$ and $\omega$.

$$F_N = \frac{W_N}{W_N^{as}} = 1 + \frac{A_1}{N - \tilde{N}} + \frac{A_2}{(N - \tilde{N})^2} + \ldots + \frac{A_K}{(N - \tilde{N})^K} + \ldots, \quad (15)$$

truncating the series, and choosing the coefficients $A_K$ from the correspondence with the known values of the coefficients $W_{L_0}$, $W_{L_0+1}$, …, $W_L$; the optimal form $W_N^{as} = ca^N N^{b-1/2}\Gamma(N + 1/2)$ [8] is used for the Lipatov asymptotics, and the parameter $\tilde{N}$ is used to analyze the uncertainty of the results. The $L_0$ value needs not to coincide with $N_0$ in Eq. (4). Indeed, the coefficient function $W_N$ continued into the complex plane has a singularity at $N = \alpha$, where $\alpha$ is the exponent of strong-coupling asymptotic expression (12) [8]. If $\alpha$ is larger than $N_0$, the interpolation with the use of all coefficients is invalid: it is necessary to set

$$W(g) = W_{N_0}g^{N_0} + \ldots + W_{N_1}g^{N_1} + \tilde{W}(g), \quad N_1 = [\alpha], \quad (16)$$

to sum the series for $\tilde{W}(g)$, and to add the contribution of the separated terms; thus, the value $[\alpha] + 1$, where square brackets means the integer part of a number, is taken for $L_0$. Analysis of the two-dimensional case [13] shows that $\alpha$ is larger than $N_0$ for almost all functions.

According to the commonly accepted tradition, in addition to the series for the functions $\beta(g)$, $\eta(g)$, and $\eta_2(g)$, we summed also the series for

$$\nu^{-1}(g) = 2 + \eta_2(g) - \eta(g), \quad \gamma^{-1}(g) = 1 - \frac{\eta_2(g)}{2 - \eta(g)}$$

in order to test the self-consistency of the results. Following [13], we allow the interpolation curves that smoothly pass through the known points, do not have significant peaks at noninteger $N$ values, rapidly approach the asymptotic behavior at large $N$ values, and have nonmonotonicity no more than the difference of the last known coefficient from the asymptotic expression.

### 3.1. Universality Class of the Ising Model ($n = 1$)

The initial information is given by the expansions [5, 7]

$$\beta(g) = -g + g^2 - 0.4224965707 g^3$$
$$+ 0.3510695978 g^4 - 0.3765268283 g^5$$
$$+ 0.49554751 g^6 - 0.749689 g^7 + \ldots$$
$$+ ca^N \Gamma(N + b) g^N + \ldots,$$

$$\eta(g) = (8/729) g^2 + 0.0009142223 g^3$$
$$+ 0.0017962229 g^4 - 0.0006536980 g^5 \quad (17)$$
$$+ 0.00138781 g^6 - 0.001697694 g^7 + \ldots$$
$$+ c' a^N \Gamma(N + b') g^N + \ldots,$$

$$\eta_2(g) = -(1/3) g + (2/27) g^2$$
$$- 0.0443102531 g^3 + 0.0395195688 g^4$$
$$- 0.0444003474 g^5 + 0.0603634414 g^6$$
$$- 0.09324948 g^7 + \ldots + c'' a^N \Gamma(N + b) g^N + \ldots$$

with the parameters

$$a = 0.14777422, \quad b = b' + 1 = 4.5,$$
$$c = 0.039962, \quad (18)$$
$$c' = 0.0017972, \quad c'' = 0.0062991$$

for the high order asymptotic expressions [4]. Some technical details of the summation procedure are discussed below.

**Function $\beta(g)$.** For the function $\beta(g)$, all interpolations with $L_0 = 1$ appeared to be unsatisfactory: interpolation curves rapidly approaching the asymptotic behavior had a sharp peak in the interval $1 < N < 2$, indicating the presence of a singularity in this interval. The estimate of strong-coupling asymptotic behavior provides $\alpha \approx 1$, confirming the presence of the singularity at $N \approx 1$ and indicating that the choice $L_0 = 2$ is correct. In this case, the interpolation curves with $\tilde{N} < -1.0$ have significant nonmonotonicity at large $N$ values, and the curves with $\tilde{N} > 1.2$ had a peak in the interval

$2 < N < 3$ (see Fig. 2); thus, the natural interpolations correspond to the interval $-1.0 < \tilde{N} < 1.2$. The summation results are shown in the inset in Fig. 2, where it is seen that

$$g^* = 1.416\text{–}1.421, \quad \omega = 0.777\text{–}0.787. \qquad (19)$$

The $g^*$ value is in agreement with the results $g^* = 1.4160 \pm 0.0015$ [5] and $g^* = 1.416 \pm 0.005$ [6] and indicates that the more recent revised result $g^* = 1.411 \pm 0.004$ [7] is doubtful.

**Function $\eta(g)$.** According to Eq. (17), the expansion of the function $\eta(g)$ begins with $g^2$. Satisfactory interpolations with $L_0 = 2$ have not been found: curves rapidly approaching the asymptotic behavior have a peak in the interval $2 < N < 3$, indicating that the exponent $\alpha$ lies in the same interval. The strong-coupling asymptotic estimate gives $\alpha \approx 2$ and indicates the necessity of the choice $L_0 = 3$. In this case, satisfactory interpolation curves are obtained only in the narrow interval $1.6 < \tilde{N} < 2.2$ (see Fig. 3); they could be treated as defective due to the presence of the peak in the interval $3 < N < 4$; however, curves of this kind ensure the exact $\eta$ value in the two-dimensional case [13]. In our opinion, such interpolations are admissible, because the amplitude of oscillations of the coefficient function is about the amplitude of oscillations of the known coefficients. The summation results are shown in the inset in Fig. 3.

**Functions $\eta_2(g)$, $\nu^{-1}(g)$, and $\gamma^{-1}(g)$.** Rough estimates of the strong coupling asymptotic behavior for the functions $\nu^{-1}(g)$ and $\eta_2(g)$ provide $\alpha \approx 2$, but they are inconsistent on the whole, breaking the relations between the functions. According to the analysis of the two-dimensional case [13], this is due to the specific properties of the function $\eta(g)$: it is small in the region $g \lesssim 10$, because the expansion coefficients are small, but it increases quite rapidly at large $g$ values. As a result, the asymptotic behavior of the functions $\nu^{-1}(g)$ and $\eta_2(g)$ contains the linear combination of the $g$ and $g^2$ contribution. It is difficult to numerically analyze this combination. For this reason, the series for $\eta_2(g)$ and $\nu^{-1}(g)$ are summed[4] at $L_0 = 3$ in order to take into account a possible singularity at $N \approx 2$, whereas the series for the function $\gamma^{-1}(g)$ is summed at $L_0 = 1$, but without the limitation of peaks at noninteger $N$ values (in view of the relation $\gamma^{-1}(g) = 1 + \eta_2(g)/[2 - \eta(g)]$, its coefficient function is expected to be regular at $N \geq 1$, but has a smeared singularity at $N \approx 1$). Figures 4–6 show the admissible interpolation curves and summation results. Table 1 presents the summation results in comparison with the results obtained using the Padé–

---

[4] The summation at $L_0 = 2$ provides the same results, but with lower uncertainty.

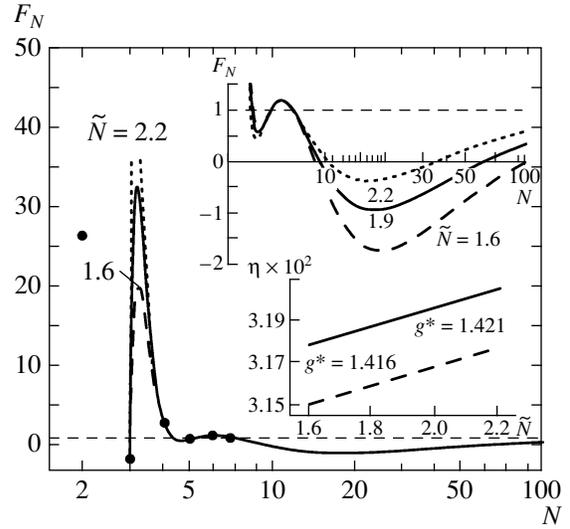

**Fig. 3.** Interpolation curves for the function $\eta(g)$ at $L_0 = 3$. The insets show the summation results at $g = g^*$.

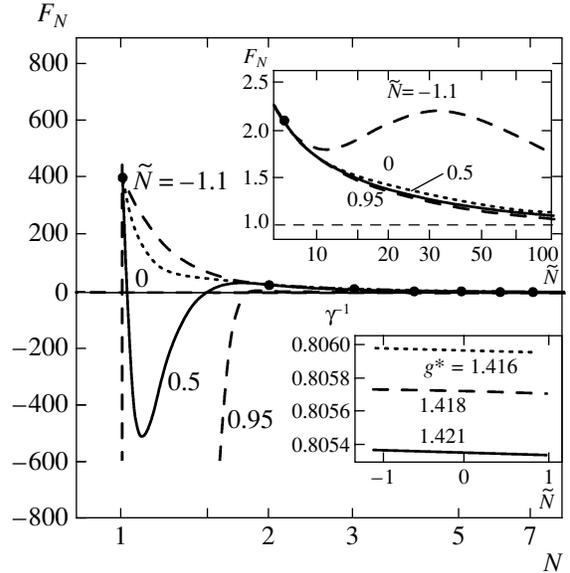

**Fig. 4.** Same as in Fig. 3, but for the function $\gamma^{-1}(g)$ at $L_0 = 1$.

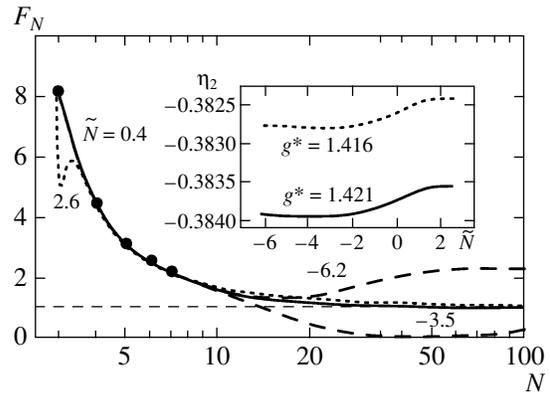

**Fig. 5.** Same as in Fig. 3, but for the function $\eta_2(g)$ at $L_0 = 3$.

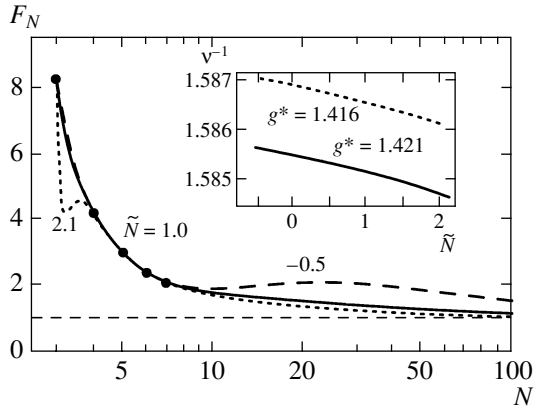

**Fig. 6.** Same as in Fig. 3, but for the function $\nu^{-1}(g)$ at $L_0 = 3$.

Borel method [5], standard conform-Borel technique [6, 7], and variational perturbation theory [14, 15].

The results for $\nu$ obtained by summing various series (using the relations $\gamma = \nu(2 - \eta)$, $\nu^{-1} = 2 + \eta_2 - \eta$, and $\nu = (1 - \gamma)/\eta_2$) are presented in Table 2. The last estimate is rather inaccurate and is not taken into account, whereas the first three estimates are very close to each other, and difference between their central values characterizes the scale of the uncontrolled systematic error

$$\delta_{syst} \approx 0.0003, \quad (20)$$

which appears because the "natural" interpolations for various functions related to each other are incompletely consistent. In the two-dimensional case [13], this effect is the main source of the error: a similar estimate provides $\delta_{syst} \approx 0.05$, which is larger than the natural summation error for most functions. In the case under consideration, $\delta_{syst}$ is rather small.

### 3.2. Universality Class of the XY Model (n = 2)

This case was discussed in detail in [16]. For completeness, we present only the final results in Table 3.

### 3.3. Universality Class of the Heisenberg Model (n = 3)

The initial information is given by the expansions [5, 7]

$$\beta(g) = -g + g^2 - 0.3832262015 g^3$$
$$+ 0.2829466813 g^4 - 0.27033330 g^5 + 0.3125559 g^6$$
$$- 0.414861 g^7 + \ldots + c a^N \Gamma(N + b) g^N + \ldots,$$

$$\eta(g) = (40/3267) g^2 + 0.0010200000 g^3$$
$$+ 0.0017919257 g^4 - 0.0005040977 g^5$$
$$+ 0.0010883237 g^6 - 0.001111499 g^7 + \ldots \quad (21)$$
$$+ c' a^N \Gamma(N + b') g^N + \ldots,$$

$$\eta_2(g) = -(5/11)g + (10/121) g^2$$
$$- 0.0525519564 g^3 + 0.039964005 g^4$$
$$- 0.0413219917 g^5 + 0.0490929344 g^6$$
$$- 0.06708630 g^7 + \ldots + c'' a^N \Gamma(N + b) g^N + \ldots$$

with the parameters [4]

$$a = 0.12090618, \quad b = b' + 1 = 5.5,$$
$$c = 0.0059609, \quad (22)$$
$$c' = 0.0003656, \quad c'' = 0.0012813.$$

The situation is qualitatively similar to the case $n = 1$. The same $L_0$ values are used, i.e., $L_0 = 1$ for $\gamma^{-1}(g)$, $L_0 = 2$ for $\beta(g)$, and $L_0 = 3$ for the other functions. The admissible interpolations correspond to the intervals $-1.0 < \tilde{N} < 1.6$ for $\beta(g)$, $1.6 < \tilde{N} < 2.3$ for $\eta(g)$, $0.4 < \tilde{N} < 2.0$ for $\nu^{-1}(g)$, $-0.6 < \tilde{N} < 2.2$ for $\eta_2(g)$, and $0.50 < \tilde{N} < 0.95$ for $\gamma^{-1}(g)$. The interpolation curves visually almost coincide with the curves shown in Figs. 2–6. The results are given in Table 4.

### 3.4. Diluted Polymers (n = 0)

The initial information is given by the expansions [5, 7]

**Table 1.** Critical exponents for the three-dimensional Ising model ($n = 1$) obtained from the field-theoretical renormalization group

| | [5] | [6] | [7] | [14] | [15] | This work |
|---|---|---|---|---|---|---|
| $\gamma$ | 1.241(4) | 1.2405(15) | 1.2396(13) | 1.241 | 1.2403(8) | 1.2411(6) |
| $\nu$ | 0.630(2) | 0.6300(15) | 0.6304(13) | 0.6305 | 0.6303(8) | 0.6306(5) |
| $\eta$ | 0.031(11) | 0.032(3) | 0.0335(25) | 0.0347(10) | 0.0335(6) | 0.0318(3) |
| $\eta_2$ | −0.382(5) | −0.3825(30) | – | – | – | −0.3832(8) |
| $\omega$ | 0.788(3) | 0.790(30) | 0.799(11) | 0.805 | 0.792(3) | 0.782(5) |
| $g^*$ | 1.4160(15) | 1.416(5) | 1.411(4) | – | – | 1.4185(25) |

$$\beta(g) = -g + g^2 - 0.4398148149 g^3$$
$$+ 0.3899226895 g^4 - 0.4473160967 g^5$$
$$+ 0.63385550 g^6 - 1.034928 g^7 + \ldots$$
$$+ c a^N \Gamma(N+b) g^N + \ldots,$$
$$\eta(g) = (1/108) g^2 + 0.0007713750 g^3$$
$$+ 0.0015898706 g^4 - 0.0006606149 g^5$$
$$+ 0.0014103421 g^6 - 0.001901867 g^7 + \ldots \quad (23)$$
$$+ c' a^N \Gamma(N+b') g^N + \ldots,$$
$$\eta_2(g) = -(1/4) g + (1/16) g^2 - 0.0357672729 g^3$$
$$+ 0.0343748465 g^4 - 0.0408958349 g^5$$
$$+ 0.0597050472 g^6 - 0.09928487 g^7 + \ldots$$
$$+ c'' a^N \Gamma(N+b) g^N + \ldots$$

with the parameters [4]
$$a = 0.16624600, \quad b = b' + 1 = 4,$$
$$c = 0.085489, \quad (24)$$
$$c' = 0.0028836, \quad c'' = 0.010107.$$

The same $L_0$ values as in the above cases are used. The admissible interpolations correspond to the intervals $-0.9 < \tilde{N} < 1.1$ for $\beta(g)$, $1.5 < \tilde{N} < 2.2$ for $\eta(g)$, $-0.7 < \tilde{N} < 2.2$ for $\nu^{-1}(g)$, $-1.5 < \tilde{N} < 2.5$ for $\eta_2(g)$, and $-0.7 < \tilde{N} < 0.95$ for $\gamma^{-1}(g)$. The results are given in Table 5.

## 4. DISCUSSION

As seen in Tables 1 and 3–5, different field-theoretical estimates are in good agreement with each other. Our results are in surprisingly good agreement with the estimates by Le Guillou and Zinn-Justin [6]: the difference between the central values is usually smaller than 0.0010 despite a quite conservative estimate of the accuracy in [6]. Such a coincidence is not accidental, since Le Guillou and Zinn-Justin [6] used interpolation to predict one or two unknown expansion coefficients and obtain some kind of an expert evaluation, but could not reduce the uncertainty of the results due to the strong dependence on the summation procedure. A more recent revision in [7] seems rather artificial and tends to take the results out their natural uncertainty. In particular, the shift of the $g^*$ values is always opposite to the shift in our calculations (see Tables 1, 3–5).[5] The agreement with the variational perturbation theory [14, 15] is also good. The positive fact is that the inclusion of information on higher orders and a more accurate estimate of the accuracy [15] approach the results for $\eta$ and $\omega$ to our values. A small discrepancy remains for the exponent $\eta$, but it is of the same magnitude as violation of the relation $\gamma = \nu(2 - \eta)$ for the central values of [15].

---
[5] We emphasize that the same information is used to calculate the function $\beta(g)$ in [6, 7] and in this work.

**Table 2.** Various estimates for the exponent $\nu$

| Series | Interval of $\nu$ | Mean value |
|---|---|---|
| $\nu^{-1}(g)$ | 0.6301–0.6311 | 0.6306 |
| $\gamma^{-1}(g), \eta(g)$ | 0.6302–0.6310 | 0.6306 |
| $\eta_2(g), \eta(g)$ | 0.6305–0.6313 | 0.6309 |
| $\gamma^{-1}(g), \eta_2(g)$ | 0.6266–0.6320 | 0.6293 |

Let us discuss the correspondence of our results with the information on the critical exponents obtained from physical experiments, Monte Carlo simulation, and high-temperature series [17].

**Case $n = 3$.** The total scatter of the Monte Carlo and high temperature results is rather large (see Fig. 7), and these results at this level are in agreement with Table 5. Our values slightly differ from the latest Monte Carlo results [18], but their difference from the most probable values following from the physical experiment is of the same order. These most probable values are located

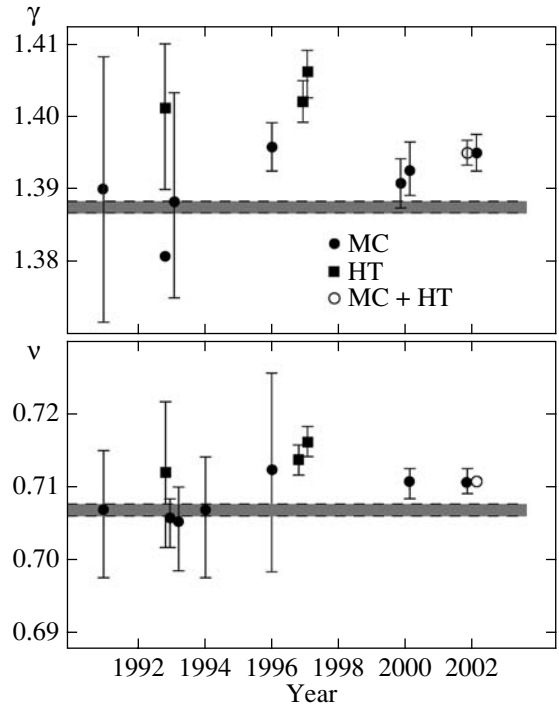

**Fig. 7.** (Points) High-temperature and Monte Carlo results for the Heisenberg model ($n = 3$) taken from [17, Table 23] in comparison with (horizontal band) the results of this work.

1126                                          POGORELOV, SUSLOV**Table 3.** Critical exponents for the XY model ($n = 2$)

|  | [5] | [6] | [7] | [14] | [15] | This work |
|---|---|---|---|---|---|---|
| $\gamma$ | 1.316(9) | 1.3160(25) | 1.3169(20) | 1.318 | 1.3164(8) | 1.3172(8) |
| $\nu$ | 0.669(3) | 0.6695(20) | 0.6703(15) | 0.6710 | 0.6704(7) | 0.6700(6) |
| $\eta$ | 0.032(15) | 0.033(4) | 0.0354(25) | 0.0356(10) | 0.0349(8) | 0.0334(2) |
| $\eta_2$ | −0.474(8) | −0.4740(25) | – | – | – | −0.4746(9) |
| $\omega$ | 0.780(10) | 0.780(25) | 0.789(11) | 0.800 | 0.784(3) | 0.778(4) |
| $g^*$ | 1.406(5) | 1.406(4) | 1.403(3) | – | – | 1.408(2) |

**Table 4.** Critical exponents for the Heisenberg model ($n = 3$)

|  | [5] | [6] | [7] | [14] | [15] | This work |
|---|---|---|---|---|---|---|
| $\gamma$ | 1.390(10) | 1.386(4) | 1.3895(50) | 1.390 | 1.3882(10) | 1.3876(9) |
| $\nu$ | 0.705(5) | 0.705(3) | 0.7073(35) | 0.7075 | 0.7062(7) | 0.7060(7) |
| $\eta$ | 0.031(22) | 0.033(4) | 0.0355(25) | 0.0350(10) | 0.0350(8) | 0.0333(3) |
| $\eta_2$ | −0.550(12) | −0.5490(35) | – | – | – | −0.5507(12) |
| $\omega$ | 0.780(20) | 0.780(20) | 0.782(13) | 0.797 | 0.783(3) | 0.778(4) |
| $g^*$ | 1.392(9) | 1.391(4) | 1.390(4) | – | – | 1.393(2) |

**Table 5.** Critical exponents for the case $n = 0$

|  | [5] | [6] | [7] | [14] | [15] | This work |
|---|---|---|---|---|---|---|
| $\gamma$ | 1.161(3) | 1.1615(20) | 1.1596(20) | 1.161 | 1.1604(8) | 1.1615(4) |
| $\nu$ | 0.588(1) | 0.5880(15) | 0.5882(11) | 0.5883 | 0.5881(8) | 0.5886(3) |
| $\eta$ | 0.026(14) | 0.027(4) | 0.0284(25) | 0.0311(10) | 0.0285(6) | 0.0272(3) |
| $\eta_2$ | −0.274(10) | −0.2745(35) | – | – | – | −0.2746(7) |
| $\omega$ | 0.794(6) | 0.800(40) | 0.812(16) | 0.810 | 0.803(3) | 0.790(6) |
| $g^*$ | 1.421(4) | 1.421(8) | 1.413(6) | – | – | 1.423(3) |

near $\gamma = 1.386$ and $\beta = 0.365$ (see Fig. 8) in good agreement with our results $\gamma = 1.3876(9)$ and $\beta = 0.3648(4)$ following from Table 5 and in worse agreement with the values $\gamma = 1.3960(10)$ and $\beta = 0.3689(3)$ presented in [18].

**Case $n = 2$.** The situation is similar to the preceding case. The total scatter of the Monte Carlo and high temperature results is rather large (see [16, Fig. 1]), but the latter results contradict Table 4 ($\gamma = 1.3178(2)$, $\nu = 0.6717(1)$, and $\eta = 0.0381(2)$ [19]) and (in the same extent) to the liquid-helium experiments, i.e., the value

$$\nu = 0.6705 \pm 0.0006, \qquad (25)$$

which is obtained from the superfluid component density measured using the data on the second-sound speed [20], and the results

$$\alpha = -0.01285 \pm 0.00038, \quad \nu = 0.67095(13) \ [21],$$
$$\alpha = -0.01056 \pm 0.00038, \quad \nu = 0.6702(1) \ [22], \qquad (26)$$
$$\alpha = -0.0127 \pm 0.0003, \quad \nu = 0.6709(1) \ [23],$$

which are obtained from the space measurements of the specific heat of helium (the relation $\alpha = 2 - d\nu$ is used).

**Case $n = 1$.** In this case, the high temperature and Monte Carlo results are numerous (see Fig. 9). The complete agreement exists for the exponent $\nu$: our estimate agrees with other field-theoretical results (see Table 3) and with almost all high-temperature and Monte Carlo data (see Fig. 9a). However, our result for $\gamma$ differs considerably from a value of 1.237–1.238 near which the high-temperature and Monte Carlo data are concentrated (see Fig. 9b).

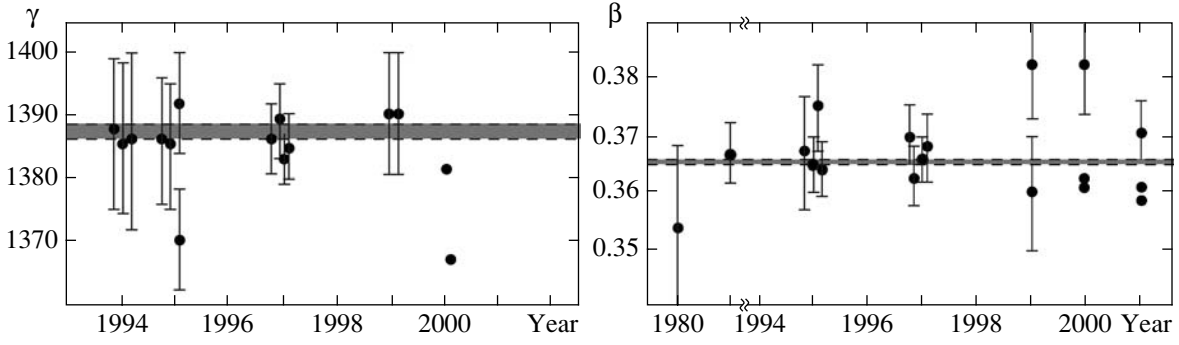

**Fig. 8.** (Points) Experimental data for the three-dimensional Heisenberg model ($n = 3$) taken from [17, Tables 24 and 25] in comparison with (horizontal band) the results of this work.

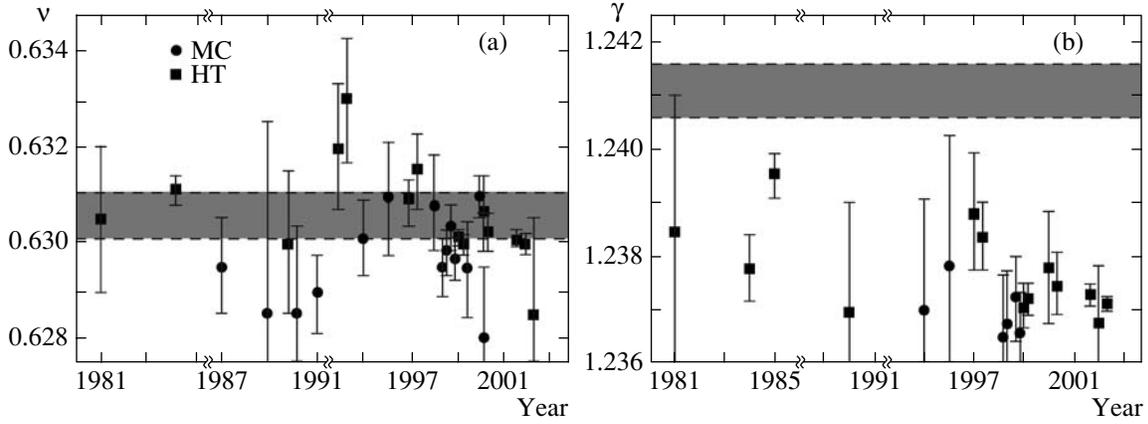

**Fig. 9.** (Points) High-temperature and Monte Carlo results for the three-dimensional Ising model ($n = 1$) taken from [17, Tables 3 and 4] in comparison with (horizontal band) the results of this work.

**Case $n = 0$.** In this case, quite accurate results for the exponent $\nu$ can be obtained by directly simulating self-avoiding random walks on a lattice. The simplicity of the algorithm allows one to collect large statistics. The latest results $\nu = 0.5876(2)$ [24], $0.5874(2)$ [25], and $0.58758(7)$ [26] slightly differ from our value presented in Table 5. The difference is not too significant and can be removed by expanding the set of the admissible interpolations, but this procedure requires the use of somewhat "unnatural" interpolation curves. The spread of the results for the exponent $\gamma$ [17] is much larger and they cannot compete with the data presented in Table 5.

## 5. DO CORRECTIONS TO THE STANDARD VALUES EXIST?

The above analysis clarifies the mathematical sense of the "standard values" (which are the central values from [6]): they correspond to the smoothest interpolation curves for the coefficient functions (see Figs. 2–6). The good agreement with other field-theoretical estimates (see Tables 1, 3–5) shows that the assumption of the extreme smoothness of the coefficient functions is also implicitly accepted in other works (see Section 2). This assumption is natural: the known expansion coefficients rapidly approach the asymptotic behavior (see Figs. 2 and 3), and a similar tendency should be expected for the subsequent coefficients. Nevertheless, there are no logical foundations for this assumption: the intermediate expansion coefficients can have an arbitrary and completely unpredictable behavior.

The reason to analyze the last possibility is the significant deviation of our $\gamma$ value for $n = 1$ from the basic high-temperature and Monte Carlo data (see Fig. 9b). This contradiction exists for all latest renormalization group estimates (see Table 1). The most suspicious in this respect is the coefficient function for $\eta(g)$ (see Fig. 4), which has an oscillating behavior. The smooth interpolation used in our work implies that such oscillations rapidly decay and do not propagate to the region of large $N$ values. However, the possibility of their comparatively slow decrease can also be considered. For example, let us set

$$F_N = F_N^{\text{reg}} + F_N^{\text{sing}}, \quad F_N^{\text{sing}} = B(-\sigma)^{-N}, \qquad (27)$$

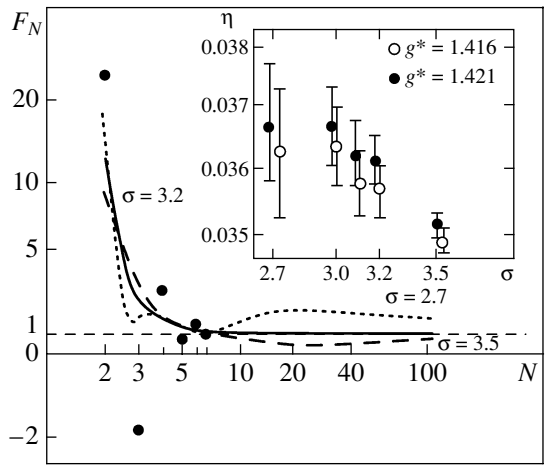

**Fig. 10.** Behavior of the regular part $F_N^{\text{reg}}$ (logarithmic scale for $F_N + 5$) introduced according to Eqs. (27) for the function $\eta(g)$. The inset is the summation results at $g = g^*$.

where $F_N^{\text{reg}}$ is described by regular expansion (15), whereas the oscillating contribution $F_N^{\text{sing}}$ has a singularity at $N = \infty$ and cannot be expanded into the series in $1/N$. Since the oscillating component is separated, extremely smooth and monotonic behavior should be expected for $F_N^{\text{reg}}$; this behavior appears in the interval $\sigma = 2.7$–$3.5$ (see Fig. 10). The singular contribution corresponds to the nonalternating series

$$W^{\text{sing}}(g) = \sum_{N=N_0}^{\infty} W_N^{\text{as}} B(-\sigma)^{-N}(-g)^N$$
$$= \sum_{N=N_0}^{\infty} cBN^{b-1/2}\Gamma\left(N + \frac{1}{2}\right)\left(\frac{ag}{\sigma}\right)^N, \quad (28)$$

which is not Borel summable: its Borel transform $B^{\text{sing}}(z)$ has a singularity at the point $z = \sigma/a$; the ambiguity of the bypass of this point generates an uncertainty of about $\exp(-\sigma/ag)$. This uncertainty is quite low, because $\sigma/ag^* \approx 15$. With the same accuracy, the sum of series (28) can be calculated by summing up to the minimum term, which will be used below. The summation results are shown in the inset in Fig. 10 and provide $\eta \approx 0.036$, which is required for the consistency of the values $\nu = 0.630$–$0.631$ and $\gamma = 1.237$–$1.238$ in view of the relation $\gamma = \nu(2 - \eta)$.

However, all singularities of the Borel transforms in the $\varphi^4$ theory lie on the negative semiaxis [10], and the perturbation-theory series are Borel summable [27]. For this reason, the oscillation decreasing law in Eqs. (27) is too slow and the above estimate should be treated as purely illustrative. Nevertheless, it shows that change in the interpretation of the oscillating contribution can provide real corrections to the standard values.

In view of the known relations, the oscillating contribution also inevitably exists for other renormalization group functions, although it is relatively small.

The above discussion shows that general investigations are necessary aimed at determining the actual law of the oscilations decrease in the coefficient functions. Without these investigations, it is impossible to analyze the existing systematic errors of the "standard values" and to increase further the accuracy of the predictions of the critical exponents.

## ACKNOWLEDGMENTS

This work was supported by the Russian Foundation for Basic Research, project no. 06-02-17541.